\documentclass[12pt,preprintnumbers]{article}
\usepackage{amsbsy,graphicx
}

\begin{document}
\vspace{10mm} \centerline{\Large \bf Non-perturbative approach
 for time-dependent}

\centerline{\Large \bf quantum mechanical system}

\vskip 1cm

 \centerline{\large Hyeong-Chan Kim$^{1}$ and Jae Hyung
Yee$^{2}$}

\vskip 1cm

\centerline{\it  Institute of Physics and Applied Physics,
 Yonsei
University} \centerline{\it Seoul 120-749, Korea}
\centerline{\it
$^{1}$hckim@phya.yonsei.ac.kr} \centerline{\it
$^{2}$jhyee@phya.yonsei.ac.kr} \vskip 2cm

\centerline{\bf Abstract} \vskip 0.5cm \noindent

We present a variational method which uses a quartic exponential function
as a trial wave-function to describe time-dependent quantum mechanical
systems. We introduce a new physical variable $y$ which is appropriate to
describe the shape of wave-packet, and calculate the effective action as a
function of both the dispersion $\sqrt{\langle \hat{q}^2\rangle}$ and $y$.
The effective potential successfully describes the transition of the
system from the false vacuum to the true vacuum. The present method well
describes the time evolution of the wave-function of the system for short
period for the quantum roll problem and describes the long-time evolution
up to 75\% accuracy. These are shown in comparison with the direct
numerical computations of wave-function. We briefly discuss the large $N$
behavior of the present approximation.

\begin{flushright}
\end{flushright}

 \vspace{1cm}
\noindent

Phase transition is one of the most important physical phenomena
in nature and has wide range of applications to condensed matter
physics, particle physics, and cosmology. Most of the studies on
this subject have been done in the framework of quasi-static
transition or using the Gaussian ansatz developed by Jackiw and
Kerman~\cite{JK}. There have been many
attempts~\cite{polly,cooper,cheetham} to go beyond the Gaussian
approximations. It is our purpose in this paper to go beyond the
Gaussian approximation in two respects. First, we need a fully
non-perturbative way which links the initial Gaussian packet (GP,
false vacuum) to the symmetry broken degenerate vacuum state (true
vacuum). Second, we try to find the relevant physical parameters
which describes the symmetry breaking effectively.

In this paper, we consider a quantum mechanical model for time-dependent
dynamics described by the potential,
\begin{eqnarray} \label{V}
V(\hat{q},t) = \frac{\lambda}{24}[\hat{q}^2 - k^2(t)]^2,
\end{eqnarray}
where $k^2(t)$ increases from a negative value to a positive number
$\kappa^2$ asymptotically. The initial GP centered at $q=0$ cannot remain
as Gaussian during the time-evolution, but evolves to the packet centered
around two minima of the potential as $k^2(t)$ approaches $\kappa^2$. For
$\kappa^2 \rightarrow \infty$, the new ground states are linear sum or
difference of two un-correlated GPs centered at each minima. In this case,
the two ground states are degenerated.

The dispersion $\langle \hat{q}^2 \rangle$ of a wavepacket may describe
the size of a GP or the distance between two packets of a double Gaussian
packet (DGP). To discern the shapes (for example, GP or DGP) of
wavepackets of the same dispersion we introduce a dimensionless quantity
$y$, which we call ``shape factor", in addition to the dispersion ($q^2$):
\begin{eqnarray} \label{exps}
q^2(t) \equiv \langle \hat{q}^2 \rangle  , ~~ y(t) \equiv
\frac{\langle \hat{q}^4 \rangle}{\langle\hat{q}^2 \rangle ^2}.
\end{eqnarray}
A similar expectation value as $y$ was calculated in Ref.~\cite{hawking}
in relation to the new inflationary scenario. To illustrate the role of
$y$ variable, consider a wavefunction which is a sum of two GPs of the
same size. If $y=1$, the density of each GP is a delta function or the two
GPs are infinitely far away so that no correlation exist between them,
which provides the lower bound of $y(\geq 1)$. If the two GPs completely
overlap, it corresponds to $y=3$, a single Gaussian packet. In between
these two states, $1< y< 3$, the two GPs are mixed and interfere with each
other. For $y>3$, there are no separable packets, and the wavefunctions
are better localized than GP~\cite{note1}.

The effective action in the variational method~\cite{JK} is given
by
\begin{eqnarray} \label{S}
\Gamma = \int dt \langle \psi, t| i \partial_t- \hat{H}
|\psi,t\rangle ,
\end{eqnarray}
where $\hat{H}=\frac{\hat{p}^2}{2m}+ V(\hat{q},t)$ and we use
$\hbar=1$. In this paper we use the trial wavefunction,
\begin{eqnarray} \label{wf:1}
\langle Q| \psi,t\rangle =
 N^{-1} \exp\left[ -
\frac{1}{2}\left(\frac{1}{2\mu^2} + i \Pi \right) Q^4
+\left(\frac{x}{\mu} +i p \right) Q^2 \right],
\end{eqnarray}
which has both of the DGP ($x\rightarrow \infty$) and the GP ($x
\rightarrow -\infty$) limits, where we assume $\mu \geq 0$. In the
static case, the double Gaussian approximation was used in
Ref.~\cite{AHK}, where a sum of two Gaussian functions is used as
a trial wave-function. However, it is difficult to generalize the
double-Gaussian method to the case for time-dependent systems. The
normalization factor $N$ can be determined by the following
integral:
\begin{eqnarray} \label{r2}
N^2=\int_{-\infty}^\infty d Q \exp\left(- \frac{Q^4}{2\mu^2}
 + \frac{2x Q^2}{\mu} \right) = \sqrt{\mu} f(x),
\end{eqnarray}
where $f(x)$ is given by~\cite{f}
\begin{eqnarray} \label{f}
f(x) = |x|^{1/2} e^{x^2}  \frac{\pi}{\sqrt{2}} \left[
I_{-\frac{1}{4}}(x^2)+ sgn(x) I_{\frac{1}{4}}(x^2) \right].
\end{eqnarray}
The dispersion and the ``shape factor" for this wavefunction are
\begin{eqnarray} \label{exps2}
q^2(t) = \frac{\mu f'}{2f} , ~~ y(t)= \frac{ 1+2x
f'/f}{{f'}^2/(2f^2)}.
\end{eqnarray}
$\displaystyle y(x)$ is a non-increasing function of $x$ from 3 to 1,
which makes the inverse function $x(y)$ be defined uniquely. We use $y$ as
a basic variable instead of $x$, because its range is bounded below by
$y=1$ for any kinds of wavepacket~\cite{note1} and it has definite
physical meaning.  The expectation values of other polynomials of
$\hat{q}^2$ can be written in terms of these parameters.

With this trial wavefunction the effective action is given by
\begin{eqnarray} \label{S1}
\Gamma =\int dt \left\{ \frac{y q^4\dot{\Pi}}{2}  - q^2\dot{p} -
2y\left[y-\frac{y-3}{Y+1}\right] \frac{q^6\Pi^2}{m} - \frac{2q^2p^2}{m}+
\frac{4 y q^4\Pi p}{m} - U(q,y) \right\},
\end{eqnarray}
where $\displaystyle Y(y)=2xf'/f=y{f'}^2/(2f^2)-1$ and the
effective potential is
\begin{eqnarray} \label{Veff}
U(q,y) = \frac{V_F(y)}{8m q^2} + \langle V \rangle ,
\end{eqnarray}
with the free potential, $V_F$, given by
\begin{eqnarray} \label{vF}
V_F(y)= 1+ \frac{(3-y)(Y+1)}{y} .
\end{eqnarray}
This free potential, coming from the expectation value $\langle
\hat{p}^2 \rangle $, represents the effect of quantum mechanical
uncertainty. The expectation value of symmetric potential,
$V(\hat{q},t)=V_0(t)+ \frac{1}{2} k(t) \hat{q}^2
+\frac{\lambda(t)}{4!} \hat{q}^4+ \frac{c(t)}{6!}
\hat{q}^6+\cdots$, with respect to $|\psi,t\rangle$ is
\begin{eqnarray} \label{Veff2}
\langle V \rangle= V_0(t) + \frac{k(t)}{2}q^2 +\frac{\lambda(t)
}{4!}y q^4
 +\frac{c(t)}{ 6!}y\left[y-\frac{y-3}{Y+1}\right] q^6 + \cdots .
\end{eqnarray}
From the action~(\ref{S1}), we notice that $\Pi$ and $p$ are
the momentum conjugates to $-yq^4/2$ and $q^2$, respectively.

Let us solve $\Pi$ and $p$ equations first:
\begin{eqnarray} \label{dy:mom}
\frac{1}{8} \frac{d}{dt} \ln (yq^4)&=& - \left[1-
\frac{y-3}{y(1+Y)} \right]
   \frac{y q^2\Pi}{m} +\frac{p}{m}, \\
- \frac{1}{4} \frac{d}{dt} \ln q^2 &=& \frac{yq^2\Pi}{m}-
\frac{p}{m}. \nonumber
\end{eqnarray}
Removing $\Pi$ and $p$ by Eq.~(\ref{dy:mom}) is just the
Legendr\'{e} transformation. Introducing new variable $\eta$ by
$\displaystyle \frac{d\eta}{dy}\equiv
D=\frac{1}{4}\sqrt{\frac{1+Y}{y(3-y)}}$, we get a quite simple
effective action in terms of $\eta$ and $q$,
\begin{eqnarray} \label{S:2}
S= \int dt \left\{\frac{m q^2}{2} \dot{\eta}^2 + \frac{m}{2}
\dot{q}^2
 - U[q,y(\eta)]  \right\}.
\end{eqnarray}
The dynamical equations of motion for $q$ and $\eta$ are given by
\begin{eqnarray} \label{eom:sb}
m\frac{d}{dt}\left[q^2 \dot{\eta}\right]+ \frac{1}{D}\left[ \frac{
V'_F(y)}{8 m q^2} +
 \partial_y\langle V \rangle \right]&=&0, \\
m \ddot{q}-\frac{ V_F}{4 m q^3}+ \partial_q\langle V\rangle &=& m
q\dot{\eta}^2 . \nonumber
\end{eqnarray}

The free potential, $\frac{V_F(y)}{8m q^2}$, has an absolute
minimum at $(y=3, ~q=\infty)$ and is positive definite. An
interesting point here is that $y=3$, the GP, actually corresponds
to $x = - \infty$. On the other hand, in the effective
potential~(\ref{vF}), $y=3$ is a regular point, which can be
extended to larger values. This property of the effective
potential implies that the trial wavefunction~(\ref{wf:1}) is
insufficient to give a full description of $y$-dependence and we
need a more general trial wavefunction for complete quantum
mechanical description which includes the range $y>3$.  The
generalization of the trial function to $|\langle Q|
\bar{\psi},t\rangle|^2 =
 \bar{N}^{-2}(1+ |z| Q^2)^{-1} |\langle Q|\psi,t\rangle|^2 $ may allow a
full description of $y$-dependence, but, is difficult to
integrate. Instead, we use a patch for the region $y>3$,
\begin{eqnarray} \label{wf:2}
\langle Q|\psi,t\rangle = N^{-1} \exp\left\{- \left(\frac{1}{4G}+
i \Pi\right) Q^2+ (-x+ i p)|Q|\right\}, ~~ y> 3,
\end{eqnarray}
where we assume $x \geq 0$ and $G\geq 0$. The wavefunction is
singular at the origin $Q=0$ due to the $|Q|$ term in the
exponent. Hence we regard the absolute value as a small $a$ limit
of $\sqrt{Q^2+ a^2}$. With this trial wavefunction, we have the
same effective action as Eq.~(\ref{S:2}) with $\displaystyle
D=\frac{\partial_y \alpha}{2\alpha \sqrt{\alpha-1}}, ~(y>3),$ and
\begin{eqnarray} \label{vF:2}
V_{F}=\left[1+2 z - \frac{2}{f}\sqrt{\frac{z}{\pi}}\right]\left(1+
\frac{2}{f} \sqrt{\frac{z}{\pi}}\right),~~ \mbox{for } y>3,
\end{eqnarray}
where $z= 2 x^2 G$, $f(z)= e^z (1- \mbox{erf}\sqrt{z})$, and
$\displaystyle \alpha= \frac{1+2 z -2 \sqrt{z}/(\sqrt{\pi}f)}{2 z
[1/(\sqrt{\pi z} f)-1]^2}$. The ``shape factor" $y(z)$
monotonically increases from 3 to 6 as a function of $z$,
\begin{eqnarray} \label{y:22}
y(z)= \frac{3+12z + 4 z^2- 2(5+ 2z)\sqrt{z}/(\sqrt{\pi}f)}{[1+2z
-2 \sqrt{z}/(\sqrt{\pi} f)]^2} .
\end{eqnarray}

As an example, let us consider the time-evolution of an initial wavepacket
given by Eq.~(\ref{wf:1}) with $(y=y_0, q=q_0)$ in the harmonic potential
$s(t) \hat{q}^2/2$. The dynamics of $y$ can be evaluated exactly from the
elliptic integral
\begin{eqnarray} \label{dY:q}
\int^{\eta}_{\eta_0}\frac{d\eta}{\sqrt{c^2-V_{F}[y(\eta)]}} = \pm
\frac{1}{2m}\int^t_{0} dt' \frac{1}{q^2(t')} ,
\end{eqnarray}
where $c^2= 4 m^2 [\dot{\eta}(0)q^2(0)]^2+ V_F[y(0)]$ is an
integration constant. The equation of motion for $q$ becomes
\begin{eqnarray} \label{largeq:eom}
\ddot{q}(t)=- s(t) q(t)+  \frac{2c^2-V_{F}(y)}{8 m^2 q^3(t)} .
\end{eqnarray}
If the system is potential free $[s(t)=0]$, $\eta(t)$
asymptotically approaches to a fixed value $\eta_f$ and $q(t)$
asymptotically increase with a constant velocity determined by the
energy conservation law. The allowed range, for constant $s(t)=s$,
of $q$ and $y$ is also determined by the energy conservation law:
\begin{eqnarray} \label{q:HO}
\frac{V_{F}(y)}{8 m q^2}+ \frac{1}{2}s q^2\leq E_{tot}= U(q_0,y_0) .
\end{eqnarray}

Let us now consider the effective potential for the classical
potential~(\ref{V}). The effective potential~(\ref{Veff})
naturally determines the true ground state with the condition,
\begin{eqnarray} \label{vac:def}
\partial_q U(q,y) =0= \partial_y U(q,y).
\end{eqnarray}
To see the behavior of the effective potential more clearly, we
variationally determine $y$, and then write down the effective
potential in $q$:
\begin{eqnarray} \label{Veff:Q}
V_{eff}(q)\equiv U(q,y_v(q)) = \frac{V_F(y_v(q))}{8m
q^2}+\frac{\lambda}{24} \left[ q^4 y_v(q)-2k^2(t) q^2 + k^4(t) \right] ,
\end{eqnarray}
where $y_v$ means that we determine $y$ by minimizing the
effective potential with the condition, $\frac{\lambda}{24}q^4 =-
\frac{V_{F}'(y_v)}{8 m q^2}$. In the static case this $y_v$ value
corresponds to the minimum position of the potential for a given
$q$, which takes $y_v(0)=3$ and $y_v(\infty) =1$. We present
figures (Fig. 1) of the effective potential~(\ref{Veff:Q}) for two
typical sets of parameters.

\begin{figure}[htbp]
\begin{center}
\includegraphics[width=1\linewidth,origin=tl]{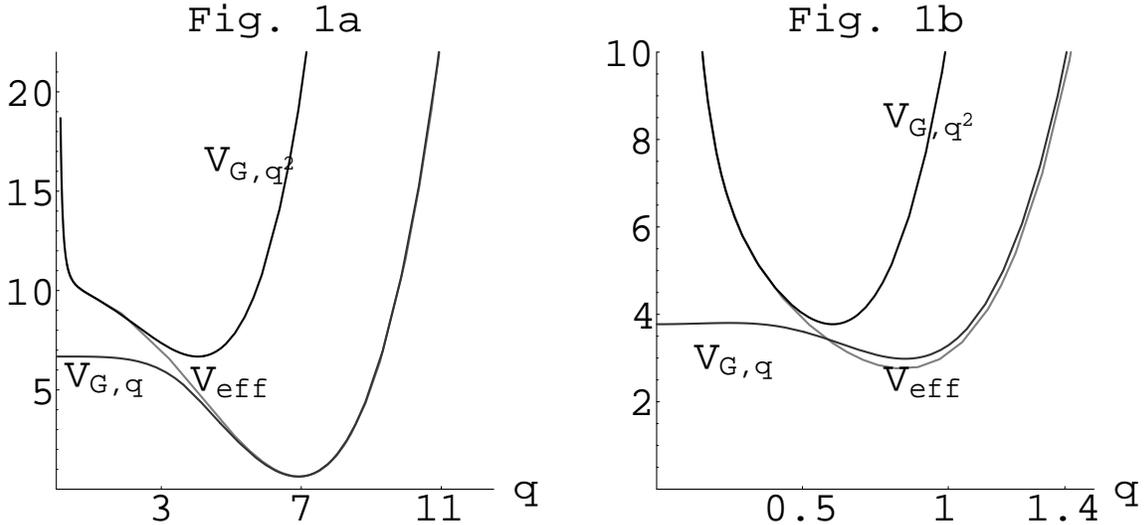}\hfill%
\end{center}
\caption{The effective potential as a function of $q$ for the
parameters $\lambda =0.1, ~ k=7,~ m=1$ (a), and $\lambda = 123,~
k=1,~ m=1 $ (b).} \label{yq:fig}
\end{figure}

In these figures, $V_{eff}=U$, $V_{G,q}$, and $V_{G,q^2}$ represent the
effective potential ~(\ref{Veff:Q}), the Gaussian approximated effective
potential for $\langle \hat{q}\rangle$, and the Gaussian approximated
effective potential for $\langle \hat{q}^2\rangle$, respectively. Here,
$V_{G,q}$ can be calculated from Eqs. (2.9) and (4.6) of
Ref.~\cite{cooper} with slight notational change ($q=\langle
\hat{Q}\rangle$) and $V_{G,q^2}$ from Eq. (2.9) of Ref.~\cite{cooper} with
$\langle \hat{Q}\rangle =0$ and $G \rightarrow q^2$. The effective
potential, $V_{eff}$, is very close to $V_{G,q^2}$ for $q <k/3$, and it
becomes close to $V_{G,q}$ for $q>2 k/3$. This clearly shows that the
initial GP is divided into a DGP, with each packet of the DGP moving as if
it is a free GP for large $k$. The value of $y_v[\sim 1+
\left(\frac{3}{\lambda m q^6} \right)^{2/5},$ for $y_v \sim 1.]$ is
effectively 1 for the most range of $q$ if $k$ is sufficiently large [$
\gg (\lambda m)^{-1/6}$], since the characteristic size of $q$ is $O(k)$.

Let us explicitly describe the dynamics of an initial GP with $q=q_0 ~(\ll
\kappa) $ for the time-dependent potential~(1). Because of the transition
(as $k(t)$ increase) $y$ eventually goes to 1 for most part of the
dynamics. The potential energy difference $\triangle V= V_{eff}(q_0,y=1)-
V_{eff}(q_0,y=3)>0$ and the presence of kinetic energy in $y$ prevent $q$
from reaching $q_0$. The time-dependence of $k(t)$ decreases the total
energy so that $q$ oscillates near the true vacuum. We present, in Fig. 2,
a solution of the differential equation~(\ref{eom:sb}) and its exact
numerical solution for the case of $k(t)$ linearly increasing to a finite
value for about a half period of $q$.
\begin{figure}[htbp]
\begin{center}
\includegraphics[width=1\linewidth,origin=tl]{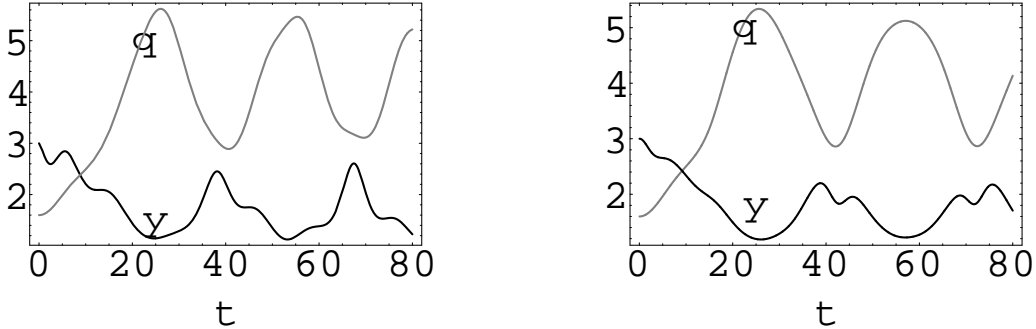}\hfill%
\end{center}
\caption{ Solution (Left) of Eq.~(\ref{eom:sb}) and exact numerical time
evolution by wave-function simulation (Right) of $q(t)$ and $y(t)$. In
this figure, we set $\lambda = 0.0123$, $m=1$, $q_0=1.6$, and $y(0)=3$.
$k(t)= t/3$ during $0\leq t \leq 15$ and remains constant afterward.}
\label{yq2:fig}
\end{figure}
In this example, we do not need the patching process by the
wavefunction~(\ref{wf:2}) since states with $y>3$ do not appear. To see
the long time behavior of the system, we present one more figure(Fig. 3).
\begin{figure}[htbp]
\begin{center}
\begin{tabular}{cc}
\includegraphics[width=.4\linewidth,origin=tl]{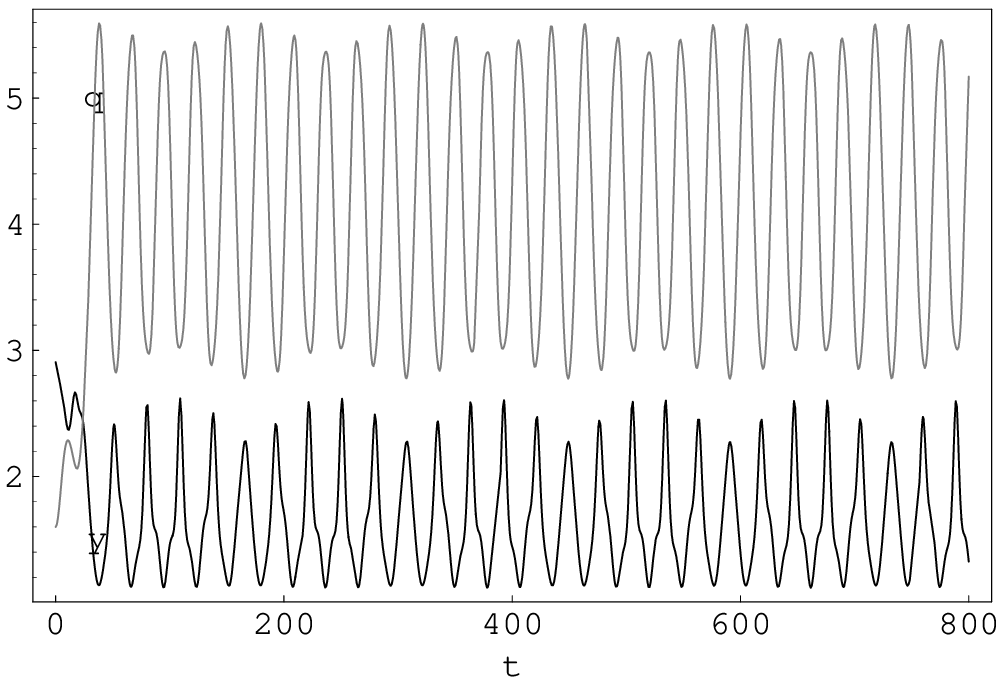}%
&\includegraphics[width=.4\linewidth,origin=tl]{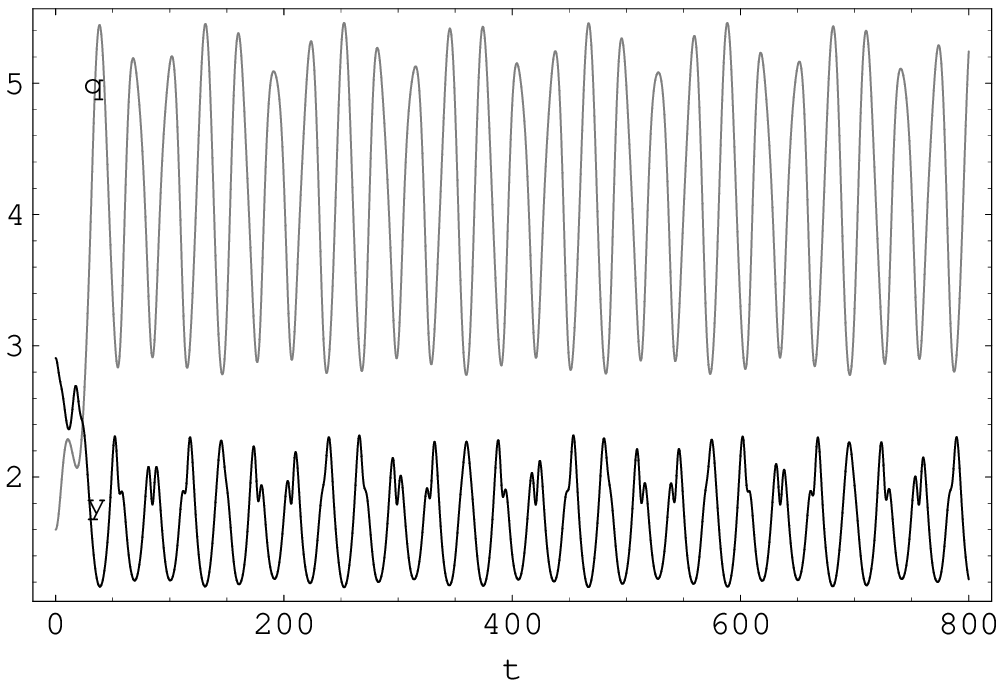}%
\end{tabular}
\end{center}
\caption{ Solution (Left) of Eq.~(\ref{eom:sb}) and exact numerical time
evolution by wave-function simulation (Right) of $q(t)$ and $y(t)$. In
this figure, we set $\lambda = 0.0123$, $m=1$, $q_0=1.6$, and
$y(0)=2.906$. $k(t)= t/3$ during $0\leq t \leq 30$ and remains constant
afterward.} \label{yq3:fig}
\end{figure}
The main characteristic feature of the long time behavior is the
oscillation of the amplitude of short time oscillation. The error of the
oscillation period in Fig. 3 is about 25\%. This error comes from the
non-exactness of the variational wave-function to the exact time-evolution
of the wave-function.

As another example, let us consider the quenched transition potential with
$k(t>0)=\kappa$ and $k(t<0)=0$. The discussions above for the
time-dependent $k(t)$ transition also applies to the present example. We
present a numerical solution of the differential equation~(\ref{eom:sb})
and its exact time evolution in Fig. 4.
\begin{figure}[htbp]
\begin{center}
\includegraphics[width=1\linewidth,origin=tl]{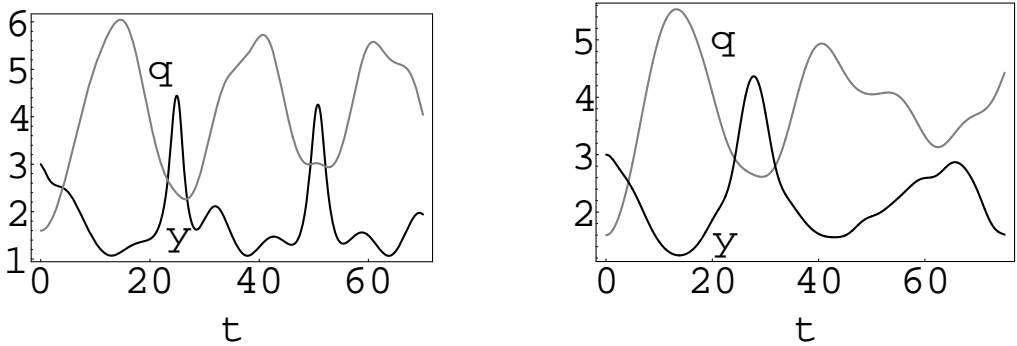}\hfill%
\end{center}
\caption{ Solution (Left) of Eq.~(\ref{eom:sb}) and exact numerical time
evolution (Right) of $q(t)$ and $y(t)$ by wave-function simulation. In
this figure, we set $\lambda = 0.0123$, $\kappa=5$, $m=1$, $q_0=1.6$, and
$y(0)=3$.} \label{qgt:fig}
\end{figure}
The state with large $y(>3)$ periodically appears. This means that we need
the patching process~(\ref{wf:2}) for the time evolution of this system.
Comparing the two results in Fig. 4, one may notice the merits and the
weakness of the present approach for the quenched transition. The present
approach explains the periodic appearance of the large ``shape factor" and
well presents the period of its occurrence, but the details of the
evolution is not exact. This discrepancy is related to the ``patched"
trial wavefunction~(\ref{wf:1}) and (\ref{wf:2}) at $y=3$. We have chosen
this artificial patching method because of its simplicity. A better
approach may be to include the excited states of (\ref{wf:1}) without
introducing the patching, (\ref{wf:2}). One of the excited states,
$\langle Q |\psi_2,t \rangle= (y-1)^{-1/2}
\left(\frac{Q^2}{q^2(t)}-1\right) \langle Q |\psi,t \rangle $, is an
orthonormal wavefunction to $\langle Q |\psi,t \rangle$. Because of the
symmetry of the potential~(\ref{V}), the odd function of $q$ cannot
contributes to the evolution. One may try the variational method by using
the following trial wavefunction:
\begin{eqnarray} \label{tf:2}
\langle Q| \bar{\psi},t\rangle =
 \bar{N}^{-1}\left[ \langle Q|\psi,t\rangle +z \langle Q |\psi_2,t
 \rangle \right],
\end{eqnarray}
where $z$ is a complex valued function of time and the normalization
factor is $\bar{N}^2= 1+ |z|^2$. This wavefunction naturally includes the
regions with $y>3$ due to the contribution of the excited state. In this
sense, the appearance of large ``shape factor" ($>3$) is the signal for
the contribution of excited states in the time evolution of the systems.
Generally, the accuracy of the approximation~(4) increases as the
potential varies slowly. We applied the present method to the case of
scalar $\phi^4$ field theory in Ref.~\cite{kim2}, and it would be
interesting to apply more realistic quantum mechanical systems in second
order phase transition.

Another point we need to speculate is the large $N$ limit. It was shown
that the large $N$ wave-function satisfies~\cite{mihaila}
\begin{eqnarray} \label{largeN}
i \frac{\partial \Phi(z,\tau)}{\partial \tau }= \left[-\frac{1}{2N^2}
\frac{\partial^2}{\partial z^2} + u(z,N)\right] \Phi(z,\tau) ,
\end{eqnarray}
where $\tau= N t$, $\displaystyle r^2 \equiv \sum_{k=1}^N x_k^2= Nz^2$,
and $u(z,N)= \frac{(N-1)(N-3)}{8N^2 z^2} + \frac{g}{8}(z^2-z_0^2)^2$. The
present approximation for symmetric state is given by
\begin{eqnarray} \label{N:dg}
\Gamma = \int d\tau \langle \psi, \tau| i \partial_\tau- \frac{1}{2N^2}
\Pi^2- u(z,N) |\psi,\tau\rangle .
\end{eqnarray}
This is the same as Eq.~(\ref{S}) with the change of parameters
$t,Q,m,V(Q,t)\rightarrow \tau,z,N^2,u(z,N)$. With the use of the trial
wave-function~(\ref{wf:1}) in~(\ref{N:dg}) the expectation value of
$\frac{1}{2N^2} \Pi^2$ is given by
\begin{eqnarray} \label{exp1:N}
2y\left[y-\frac{y-3}{Y+1}\right] \frac{q^6\Pi^2}{N^2} +
\frac{2q^2p^2}{N^2}- \frac{4 y q^4\Pi p}{N^2} + \frac{V_{F}(y)}{8N^2},
\end{eqnarray}
where the first three terms in Eq.~(\ref{exp1:N}) are $O(1)$, and in the
large $N$ limit the quantum mechanical effects on the potential
$\frac{V_{F}(y)}{8N^2}$ effectively vanishes as $O(1/N^2)$. In the absence
of this quantum mechanical term, the equations of motion in the large $N$
limits for the present quartic exponential approximation with $y=1$ is the
same as that of the Gaussian approximation centered at $z \neq 0$. Since
the Gaussian approximation was proven to be the same as large $N$
approximation~\cite{cooper2}, the present approximation is equivalent to
the large $N$ approximation for $N \rightarrow \infty$.

\vspace{0.5cm}
~\\
{\Large {\bf Acknowledgments}} \\
~\\
This work was supported in part by Korea Research Foundation under Project
number KRF-2001-005-D2003 (H.-C.K. and J.H.Y.).



\begin{thebibliography}{99}
\bibitem{note1}
$y$ has no upper limit. As an example, we consider $\displaystyle
|\langle Q|\psi,t\rangle|^2 = \frac{e^{-Q^2/G}}{N^2(1+ x Q^2)}$.
The shape factor for this packet is $\displaystyle
y=\frac{1-(1-1/(2\alpha))\bar{g}}{(1-\bar{g})^2}$, where
$\alpha=\sqrt{Gx}$, and $\displaystyle \bar{g}=
\sqrt{\frac{1}{\pi\alpha}}
\left\{e^{\alpha}[1-\mbox{erf}(\sqrt{\alpha})]\right\}^{-1}$.
\end{thebibliography}
\end{document}